\documentclass[twocolumn,showpacs,preprintnumbers,amsmath,amssymb]{revtex4}
\usepackage{graphicx}
\begin{document}
\title {Luttinger liquid with strong spin-orbital coupling
and Zeeman splitting in quantum wires}
\author{Yue Yu$^1$,  Yuchuan Wen$^1$, Jinbin Li$^1$, Zhaobin Su$^1$ and S. T. Chui$^2$
 }
\affiliation{1. Institute of Theoretical Physics, Chinese Academy
of Sciences, P.O. Box 2735, Beijing 100080, China\\
 2. Bartol Research Institute, University of Delaware,
Newark, Delaware 17916}
\date{\today}
\begin{abstract}
We study a one-dimensional interacting electron gas with the
strong Rashba spin-orbit coupling and Zeeman splitting in a
quantum well. A bosonization theory is developed for this system.
The tunneling current may deviate from a simple power law which is
that in an ordinary Luttinger liquid. The microscopic interacting
coupling and the spin-orbital parameter may be measured by varying
the external magnetic field in the tunneling experiment.
\end{abstract}


\maketitle

The realization of the quasi-one-dimensional conductor in quantum
wires, carbon nanotubes and DNA molecules provided possible
realization of Luttinger liquid systems \cite{Hald}. Quantum wires
fabricated from narrow gap semiconductors, e.g., InAs, are highly
interesting objects in the spintronics due to their rich spin
transport properties and their application as spin transistors.

The Rashba effect \cite{Rash}, arising from the confining
potential which is necessary to fabricate the quantum wires,
dominates the behavior of the spin-orbit(SO) coupling in narrow
gap semiconductors with heavy and light hole bands. Moroz and
Barnes have shown that in a strong Rashba spin-orbital coupling,
the dispersion of the electron in the quasi-one dimensional
quantum well is drastically deformed \cite{mon1}. The left and
right Fermi velocities are not equal for a spin-$s$ electron while
the dispersions of the spin-up and down electrons have a mirror
symmetry. Furthermore, a bosonization theory including such a
strong Rashba spin-orbit coupling has been constructed\cite{mor}.
It was found that, instead of the separated spin and charge
excitations, two new branches of excitations with mixed spin and
charge were found and the characteristic of the Luttinger liquid
are modified by the SO coupling parameter \cite{egg}.

To detect the SO coupling, an external magnetic field was applied
\cite{shu}. Moreover, reliable techniques to measure the SO
parameter need to be further developed. On the other hand, the
Zeeman splitting due to the external magnetic field also modifies
the transport behavior of the Luttinger liquid \cite{Kim}. A
further question raised here is what will happen if both the SO
interaction and Zeeman splitting are considered simultaneously. We
find that a bosonization technique combining the methods used in
Moroz et al \cite{mor} and Kim et al \cite{Kim} can be employed to
solve the present problem. It is seen that there are four
independent chiral excitations with different sound velocities. In
an interacting electron system, this leads to different power laws
for the left and right moving single particle densities of states
(DOSs) as a function of the energy from the Fermi level. The
tunneling current between a 3-dimensional metal(say, a STM tip)
and the quantum wire may deviate from a simple power law and it
may measure this exponent asymmetry in DOSs, and then the
microscopic interacting coupling and the Rashba parameter.

Moroz and Barnes have computed the band structure of this model
with an external magnetic field \cite{mon1}. The non-interacting
electron Hamiltonian under consideration reads
\begin{eqnarray}
H_0&=&\frac{1}{2m_b}(\hat{\bf p}+{e\over c}{\bf A})^2+V({\bf r})+
H_Z+H_{SO},\label{ham}\\
H_Z&=&-\frac{g}{2}\mu_BB\sigma_z\nonumber
\end{eqnarray}
where $m_b$ is the electron band mass, $g$ is the Lande $g$
factor, and $\mu_B$ is the Bohr magneton. The vector potential
${\bf A}=-Bx\hat y$ which gives the magnetic field $B\hat
z=\nabla\times {\bf A}$. The transverse confining potential is
approximated by a parabola. The SO Hamiltonian is given by
\begin{eqnarray}
H_{SO}=\frac{\alpha}{\hbar}[{\vec \sigma}\times(\hat {\bf
p}+\frac{e}{c}{\bf A})]_z,
\end{eqnarray}
where $\alpha$ takes $10^{-10}\sim 10^{-9}$ eVcm \cite{Rash}. The
strong Rashba effect appears if $l_\omega/l_\alpha\geq 1.4$ for
$l_\alpha=\hbar^2/2m_b\alpha$ and
$l_\omega=\sqrt{\hbar/m_b\omega}$ where $\omega$ is the frequency
of the confining potential. In this case, Moroz et al found the
linearized second quantized Hamiltonian for $B=0$ can be written
as \cite{mor}
\begin{eqnarray}
H_0(B=0)&=&\sum_k v_1k(c^\dagger_{k,R,+}c_{k,R,+}
-c^\dagger_{k,L,-}c_{k,L,-})\nonumber\\
&+&v_2k(c^\dagger_{k,R,-}c_{k,R,-} -c^\dagger_{k,L,+}c_{k,L,+}),
\label{be0}
\end{eqnarray}
where the Fermi velocities $v_{1,2}$ are dependent on spin and
chirality, which can be obtained by calculating the spectrum of
the system. The difference between the velocities, $\delta
v_F(\alpha,\omega)=v_1-v_2$ monotonically increases as $\alpha$ is
enhanced. The electron fields $c_{k,\gamma,s}$ for $\gamma=R,L$
and $s=\pm$ relate to $c_{k,\gamma,\delta}$
($\delta=\downarrow,\uparrow$) by $c_{k,\gamma,\delta}=a_{\delta
s}(\alpha,\omega,k)c_{k,\gamma,s}$, where $a_{\delta
s}(\alpha,\omega,k)$ is a spin rotation matrix which can be
obtained by numerically solving the single particle Schr\"odinger
equation corresponding to eq.(\ref{ham}) \cite{mon1}. It may be
expanded as $a_{\delta s}=\delta_{\delta,s}+O(\delta v_F/v_F)$
with the average Fermi velocity $v_F=(v_1+v_2)/2$.

For a weak magnetic field, its coupling to the orbital can be
neglected if we consider the low-lying excitation. In the
following, we only keep the Zeeman term with respect to the
magnetic field. Define $k_F=m_b v_F$ and $\delta v_B=g\mu_B
B/k_F$, the Zeeman term reads $H_Z=2\delta v_Bk_F\sum (\sigma_z)_
{\delta\delta'}a^{*-1}_{\delta s}a^{-1}_{\delta'
s'}c^\dagger_{k,s}c_{k,s'}\approx 2\delta v_Bk_F\sum (\sigma_z)
_{ss'}c^\dagger_{k,s}c_{k,s'}+O(\delta v_B\delta v_F/k_F^2)$.
Thus, for $\delta v_F\sim \delta v_B\ll v_F$, the Zeeman term can
be directly added to the Hamiltonian (\ref{be0}). The dispersion
of the lowest subband now has a Zeeman splitting as sketched in
Fig. 1. The linearized non-interacting electron Hamiltonian is
given by
\begin{eqnarray}
H_0=\sum_{k,\gamma,s} v^s_\gamma
kc^\dagger_{k,\gamma,s}c_{k,\gamma,s},
\end{eqnarray}
where $v^s_\gamma=\pm (v_F+s\delta v_F/2)+s\delta v_B/2$ are four
different sound velocities.\\

{\noindent \it Bosonized Hamiltonian: } Using the above known
result, we are going to the bosonized formalism. We take the
electron interaction to be the SU(2) invariant form $H_{int}=2\pi
U\int dy \psi^\dagger_\delta(y)\psi_\delta(y)
\psi^\dagger_{\delta'}(y)\psi_{\delta'}(y)$ for the electron
field. At low energies, the interaction can be divided into the
forward, backwordard, umklapp and oscillating terms. For the
semiconduct quantum well, it is enough to only keep the forward
scattering in low energies \cite{CP}. In the bosonization theory,
the electron operators $\psi_{\gamma,l}(x)$ are defined by
$\psi_{\gamma,l}(x)\propto
e^{-i(-1)^l2\sqrt{\pi}\phi_{\gamma,l}(x)}$. Assuming $q$ to be the
low-lying excitation wave vector and $q\ll\delta v_F\sim \delta
v_B\ll v_F$, the bosonized form of the whole Hamiltonian is easily
arrived at:
\begin{eqnarray}
H&=&\frac{1}2\int dx\biggr[v_\rho K_\rho(\partial_x\theta_\rho)^2
+\frac{v_\rho}{K_\rho}(\partial_x\phi_\rho)^2\nonumber\\
&+&v_\sigma K_\sigma(\partial_x\theta_\sigma)^2
+\frac{v_\sigma}{K_\sigma}(\partial_x\phi_\sigma)^2\nonumber\\
&+&\delta v_B(\partial_x \phi_\rho\partial_x\phi_\sigma+\partial_x
\theta_\rho\partial_x\theta_\sigma)\nonumber\\
&+&\delta v_F(\partial_x \phi_\rho\partial_x\theta_\sigma+
\partial_x \phi_\sigma\partial_x\theta_\rho)\biggr], \label{boh}
\end{eqnarray}
where the subindices $\rho,\sigma$ are the charge and $s$-spin
degrees of freedom; $v_{\rho,\sigma}=[(v\pm U)^2-U^2]^{1/2}$ and
$K_{\rho,\sigma}=\sqrt{v/(v\pm 2U)}$. Formally, the bosonized
Hamiltonian (\ref{boh}) is a direct generalization of the theory
given in \cite{mor} or \cite{Kim}. However, the physics included
in such a generalization is richer.\\

{\noindent \it Excitaions: } After diagonalizing the Hamiltonian
(\ref{boh}), the theory cannot be written as that of a standard
harmonic fluid form of the Luttinger liquid \cite{Hald1}. We have
four branches of the chiral excitations with the velocities
$u_{\gamma}^{l}$, which are the absolute value of the solutions of
the eigen equations. When $\delta v=0$ or $\alpha=0$, one gets
back the cases discussed in Refs.\cite{mor} and \cite{Kim}: There
are two excitations whose velocities are
$u_{\pm}^{R}=u_{\pm}^L=[v^2+(\delta v_F/2)^2 \pm
2v\sqrt{U^2+(\delta v_F/2)^2}]^{1/2}$ or $[v^2+(\delta v_B/2)^2\pm
2\sqrt{v^2(\delta v_B/2)^2+v_1v_2U^2}]^{1/2}$, respectively. Using
the sound velocities $u_\gamma^l$, one can solve the eigenvectors
$\beta_a^{(\gamma,l)}$ which obey the sympletic orthogonal
condition: $\sum_a
(-1)^{a+1}\beta_a^{(\gamma,l)}\beta_a^{(\gamma',l')}=(-1)^{1+l}\delta_{\gamma,\gamma'}\delta_{l,l'}.$
The old fields $\phi_{\rho,\sigma}^{R,L}(x)$ can be expressed in
terms of the four new chiral fields $\Phi_\gamma^l(x)$ with
velocities $u_\gamma^l$: $
\phi_\rho^R=\beta_1^{(+,R)}\Phi_+^R+\beta_2^{(+,R)}\Phi_+^L+\beta_3^{(+,R)}\Phi_{-}^R+
\beta_4^{(+,R)}\Phi_{-}^L$ and $
\phi_\sigma^L=\beta_1^{(-,L)}\Phi_{+}^R+\beta_2^{(-,L)}\Phi_+^L+\beta_3^{(-,L)}\Phi_{-}^R+
\beta_4^{(-,L)}\Phi_{-}^L$ and so on. In these new fields, the
Hamiltonian (\ref{boh}) is diagonalized.\\

{\noindent \it Single particle correlation functions and tunneling
current:} The electron correlation function is a function of both
the SO coupling and the external magnetic field. We first
calculate the single particle correlation function , the Fourier
transform at $x=0$ of which corresponds to the single particle
DOS. Using the relation between the old and new fields, the
$s(=\pm)$-dependent correlation functions are given by
\begin{eqnarray}
&&G_{+,R}(x,t)\sim x_{+,R}^{-(\beta^{R}_1)^2}
x_{+,L}^{-(\beta^{R}_2)^2}x_{-,R}^{-(\beta^{R}_3)^2}
x_{-,L}^{-(\beta^{R}_4)^2},\\
&&G_{-,R}(x,t)\sim x_{+,R}^{-(\delta\beta^R_1)^2},
x_{+,L}^{-(\delta\beta^R_2)^2}x_{-,R}^{-(\delta\beta^R_3)^2}
x_{-,L}^{-(\delta\beta^R_4)^2},\nonumber
\end{eqnarray}
where $\beta^R_a=\frac{1}{\sqrt 2}(\beta^{(+R)}_a+\beta_a^{(-R)})$
and $\delta\beta^R_a=\frac{1}{\sqrt
2}(\beta^{(+R)}_a-\beta_a^{(-R)})$. The spin($\delta$) dependent
correlation functions are given by
\begin{eqnarray}
&&G_{\uparrow,R}=|a_{\uparrow,+}|^2G_{+,R}(x,t)+
|a_{\uparrow,-}|^2
G_{-,R}(x,t),\\
&&G_{\downarrow,R}=|a_{\downarrow,+}|^2G_{+,R}(x,t)+|a_{\downarrow,-}|^2
G_{-,R}(x,t). \nonumber
\end{eqnarray}
Furthermore, since the spin rotational invariance, one has
$G_{\rho,R}(x,t)=\sum_\delta G_{\delta,R}(x,t)=\sum_s
G_{s,R}(x,t)$. The right-moving single particle DOS, which is
given by the Fourier transform of $G_{\rho,R}(x,t)$ at $x=0$,
reads
\begin{eqnarray}
n_R(\omega)\sim A_R\omega^{\beta_R}
\end{eqnarray}
where $\beta_R={\rm min}\{ \sum_a(\delta\beta_a^{R})^2-1,
\sum_a(\beta_a^{R})^2-1\}$. Similarly, one can calculate
$n_L(\omega)\sim A_L\omega^{\beta_L}$ with $\beta_L={\rm min}\{
\sum_a(\delta\beta_a^{L})^2-1,\sum_a(\beta_a^{L})^2-1\}$. In Fig.
2, we show the exponents $\beta_l$ as functions of $\delta v_B$
and $\delta v_F$. $\beta_R\ne\beta_L$ for all non-vanishing
$\delta v_F$ and $\delta v_B$. This is a new feature which was not
observed before. The chiral DOS is an experimental observable. For
example, in a tunneling current measurement between a
3-dimensional metal (say, an STM tip) and a quantum wire, the
tunneling current is given by Fermi's golden rule:
\begin{eqnarray}
I\propto \int d\varepsilon
[n^{(3)}_R(\varepsilon)n_R(\varepsilon-eV)
-n^{(3)}_L(\varepsilon-eV)n_L(\varepsilon)],
\end{eqnarray}
where $n^{(3)}_l$ is the single particle DOS of the Fermi liquid.
At zero energies, the differential conductance is given by
\begin{eqnarray}
\frac{dI}{dV}\propto n_L(eV)+\cos(\beta_R\pi) n_R(eV).
\end{eqnarray}
For an ordinary Luttinger liquid, $\beta_R=\beta_L$ and
$dI/dV\propto n(eV)$, which has been recognized by Matveev and
Glazman\cite{exp1}. Now, due to $\beta_R\ne\beta_L$ while they are
pretty close, the differential conductance is not a simple power
law. Such an asymmetry of the exponents in DOSs and the complex in
the differential conductance are a unique manifestation of the
co-ordination between the Zeeman splitting and the SO precession
in a Luttinger liquid. Furthermore, this differential conductance
and the exponents measurements also imply that the interacting
coupling $U$ and the SO parameter $\alpha$ can be determined
through the measured values of $\beta_l^2$ as a function of the
external magnetic field. This may provide a reliable way to
determine the SO parameter.\\

{\noindent \it Density-density correlations: } The density-density
correlation functions describe the density fluctuations of the
system. The charge and $s$-spin density fluctuations are given by
\begin{eqnarray}
&&\langle0|\rho_{\rho,\sigma}(x,0)\rho_{\rho,\sigma}(0,0)|0\rangle\nonumber\\
&&\sim -\frac{K'_{\rho,\sigma}}{2\pi^2x^2}+{\rm
const}\cdot\frac{\cos(4\pi k_Fx)}{x^{2K'_{\rho,\sigma}}},
\end{eqnarray}
where $K'_\rho=(\beta^{(+R)}_1+\beta^{(+L)}_1)^2
+(\beta^{(+R)}_3+\beta^{(+L)}_3)^2
=(\beta^{(+R)}_2+\beta^{(+L)}_2)^2+(\beta^{(+R)}_4+\beta^{(+L)}_4)^2$
and $K'_\sigma=(\beta^{(-R)}_1+\beta^{(-L)}_1)^2+(\beta^{(-R)}_3+
\beta^{(-L)}_3)^2 =(\beta^{(-R)}_2+\beta^{(-L)}_2)^2
+(\beta^{(-R)}_4+\beta^{(-L)}_4)^2$.

For$S^y=\frac{1}2\psi^\dagger_s\sigma^y_{ss'} \psi_{s'}$, the
$y$-component of the $s$-spin operator, the correlation function
is given by
\begin{eqnarray}
\langle 0|S^y (x,0)S^y (0,0)|0\rangle&\sim& \cos(4\pi
k_Fx)x^{-K'_\sigma/2}\nonumber\\
&+&{\rm const}\cdot x^{-2-(\frac{1}{2\sqrt K'_\sigma}-{\sqrt
K'_\sigma})^2}
\end{eqnarray}
\\

{\noindent \it Transport in an infinitely long pure wire : }The
electron transport in the Luttinger liquid is affected by the SO
coupling and the external magnetic field. From the continuity
equation, the current operator is defined by
$j(x,\tau)=-\frac{i}{\sqrt
\pi}\partial_\tau(\phi_{\rho,R}+\phi_{\rho,L})$ where $\tau=it$.
Here we only show the calculation of the conductance for a pure
infinitely long quantum wire. It is also straightforward to
calculate the conductance with impurity scatterings\cite{chui},
although we shall not show it here. The conductivity is defined by
\begin{eqnarray}
\sigma_\omega(x)=\frac{e^2}{\pi\omega}\int_0^{1/T}d\tau\langle
0|T_\tau\partial_\tau\phi_\rho(x,\tau)\partial_\tau\phi_\rho(0,0)|0
\rangle e^{-i\omega \tau}.
\end{eqnarray}
Calculating $\langle 0|T_\tau\phi_\rho(x,\tau)\phi_\rho(0,0)|0
\rangle$, one has the conductance $ G=2K'_\rho\frac{e^2}h. $

Similar to the charge current, we can define the $s$-spin current
$j_\sigma(x,\tau)=-\frac{i}{\sqrt\pi}\partial_\tau(\phi_{\sigma,R}+\phi_{\sigma,L})$.
The same calculation gives rise to the $s$-spin conductance $
G_\sigma=2K'_\sigma\frac{e^2}{h}$. The $\delta$ spin current are
given by $ j_{\rm spin}(x,\tau)=(|a_{\uparrow +}|^2-|a_{\downarrow
+}|^2) j_\sigma(x,\tau)$. The $\delta$ spin conductance in the low
energy limit is given by $ G_{\rm spin}=2(|a_{\uparrow
+}|^2-|a_{\downarrow +}|^2)K_{\rm spin} \frac{e^2}h$ and $K_{\rm
spin}=K'_\sigma$. The consequences arising from the above
discussion are (i) For $\alpha=B=0$, $j_{\rm spin}=j_\sigma$ and
$G_{\rm spin}=G_\sigma\ne 0$ because of the SU(2) invariance of
the system. (ii) $j_{\rm spin}$ and $G_{\rm spin}$ portray the
$B$-dependence of the system while $j_\sigma$ and $G_\sigma$
portray the dependence of the system on the SO coupling. For
$\alpha\ne 0$ and $B=0$, the spin conductivities
$G_\uparrow=G_\downarrow$, i.e., $G_{\rm spin}=0$ if $B=0$. When
$U=0$, this agrees with previous work\cite{mol}. For $\alpha=0$,
$j_{\rm spin}=j_\sigma$. The SO effect is reflected by $G_\sigma$
which increases as $U$ goes up. This enhancement of the SO effect
has been observed for a two-dimensional electron gas \cite{chen}.
(iii) When $B>0$ and $\alpha>0$, $j_{\rm spin}< j_\sigma$. On the
other hand, $G_{\rm spin}$ grows as the interaction becomes
stronger because $G_\sigma$ grows. (iv) For given $\alpha$ ($B$),
the magneto-polarization is given by
\begin{eqnarray}
P&=&\frac{j_\uparrow-j_\downarrow}{j_\uparrow+j_\downarrow}
=(|a_{\uparrow +}|^2-|a_{\downarrow
+}|^2)\frac{K'_\sigma}{K'_\rho}\nonumber\\
&=&C\cdot\frac{\delta v_B}{v_F}\cdot \frac{K'_\sigma}{K'_\rho},
\end{eqnarray}
where $a_{\delta s}$ is a function of $\alpha,\omega,k_F$ and $B$;
$C$ is constant with order one. It is seen that the interaction
can strongly enhance the polarization. The detailed
numerical analysis will be given in further works.\\

In conclusion, we have built a low energy effective theory when
the Zeeman splitting and strong Rashba SO coupling in a Luttinger
liquid are turned on simultaneously. Different from that in the
ordinary two component Luttinger liquid, four chiral excitations
with different sound velocities were found. An asymmetry of the
single particle DOS has been found, which implies a deviate from
the single power law of the tunneling current between a
3-dimensional metal and the quantum wire. The variation of the
differential conductance as a function of the external magnetic
field also allowed for the determination of the interacting
coupling and the SO parameter. The concrete numerical results will
be given elsewhere.

This work was supported in part by Chinese NSF and US NSF.

{\noindent Fig. 1 The dispersion with both strong Rashba SO
coupling and weak Zeeman splitting.}

{\noindent Fig. 2 (a) $\beta_l$ as a function of the normalized
magnetic field $\delta v_B/2=g\mu_B/2k_F$. (b) $\beta_l$ as a
function of the spin orbit coupling strength $\delta v_F$. The
circle is for $\alpha=0$ ($\beta_R=\beta_L$). The filled(empty)
up-triangle is for $\beta_R(\beta_L)$ $\delta v_F/v_F=0.2$(a),
$\delta v_B/v_F=1/3$ (b). The filled(empty) down-triangle is for
$\beta_R(\beta_L)$ with $\delta v_F/v_F=0.4$ (a), $\delta
v_B/v_F=2/3$ (b). The interaction strength $U=0.25v_F$.}

\end{document}